\documentclass[aps,prl,floatfix,twocolumn,reprint,amsmath,amssymb,superscriptaddress]{revtex4-1}
\usepackage{amsfonts}
\usepackage{mathrsfs}
\usepackage{amsmath}
\usepackage{color}
\usepackage{natbib}
\usepackage{graphicx}
\usepackage{bm}
\usepackage{amssymb}
\usepackage{xspace}
\usepackage{epstopdf}
\usepackage{dcolumn}
\usepackage{longtable}
\usepackage{multirow}
\usepackage[colorlinks=true, letterpaper=true, pdfstartview=FitV, linkcolor=blue, citecolor=blue, urlcolor=blue]{hyperref}
\UseRawInputEncoding

\makeatletter

\newcommand{\Rmnum}[1]{\expandafter\@slowromancap\romannumeral #1@}
\makeatother

\begin{document}

\title{Flat-band and multi-dimensional fermions in Pb$_{10}$(PO$_{4}$)$_{6}$O$_{4}$}
\date{\today}

\author{Botao Fu}
\affiliation{College of Physics and Electronic Engineering, Center for Computational Sciences, Sichuan Normal University, Chengdu, 610068, China}
\author{Qin He}
\affiliation{College of Physics and Electronic Engineering, Center for Computational Sciences, Sichuan Normal University, Chengdu, 610068, China}

\author{Xiao-Ping Li}
\email{xpli@imu.edu.cn}
\affiliation{School of Physical Science and Technology, Inner Mongolia University, Hohhot 010021, China}

\begin{abstract}
\setlength{\parindent}{2em}
Employing a combination of first-principles calculations and low-energy effective models, we present a comprehensive investigation on the electronic structure of Pb$_{10}$(PO$_{4}$)$_{6}$O$_{4}$, which exhibits remarkable quasi-one-dimensional flat-band around the Fermi level that contains novel multi-dimensional fermions. These flat bands predominantly originate from $p_x/p_y$ orbital of the oxygen molecules chain at $4e$ Wyckoff positions, and thus can be well-captured by a four-band tight-binding model. Furthermore, the abundant crystal symmetry inherent to Pb$_{10}$(PO$_{4}$)$_{6}$O$_{4}$ provides an ideal platform for the emergence of various multi-dimensional fermions, including a 0D four-fold degenerated Dirac fermion with quadratic dispersion, a 1D quadratic/linear nodal-line (QNL/LNL) fermion along symmetric $k$-paths, 1D hourglass nodal-line (HNL) fermion linked to the Dirac fermion, and a 2D symmetry-enforced nodal surface (NS) found on the $k_z$=$\pi$ plane. Moreover, when considering the weak ferromagnetic order, Pb$_{10}$(PO$_{4}$)$_{6}$O$_{4}$ transforms into a rare semi-half-metal, which is characterized by the presence of Dirac fermion and HNL fermion at the Fermi level for a single spin channel exhibiting 100$\%$ spin polarization.
Our findings reveal the coexistence of flat bands, diverse topological semimetal states and ferromagnetism within in Pb$_{10}$(PO$_{4}$)$_{6}$O$_{4}$,  which may provide valuable insights for further exploring intriguing interplay between superconductivity and exotic electronic states.

\end{abstract}

\maketitle

\section{Introduction}
Recently, attention has turned to a new class of materials with unique structures, exemplified by the groundbreaking discovery of Cu-doped lead-apatite (LK-99), Pb$_{10}$(PO$_{4}$)$_{6}$O, which has been reported as the first room-temperature ambient-pressure superconductor\cite{lee2023roomtemperature,lee2023superconductor}.
Unfortunately, as more precise experimental observations accumulate\cite{garisto2023lk}, the mystery surrounding superconductivity in LK-99 is gradually unraveling, and researchers are increasingly inclined to refute the notion of room-temperature superconductivity in LK-99. Nevertheless, this system is very interesting in its own electronic structure. The structure is based on the P63/m-Pb$_{10}$(PO$_{4}$)$_{6}$O structure model as refined by Krivovichev and Burns in 2003\cite{krivovichev2003crystal}. The Apatites\cite{griffin2023origin} belongs to a group of materials with the general formula A$_{10}$(TO$_{4}$)$_6$X$_{2+x}$, where A is the alkaline or rare earth metal, T=Ge, Si, or P, and X=halide, O or OH.
The parent material, Pb$_{10}$(PO$_{4}$)$_{6}$O, can be modulated through various factors, such as element substitutions\cite{liu2011synthetic,baikie2014influence,fleet2010structural} and concentration adjustments\cite{ACSEnergyLetters2023}. Consequently, the investigation of derivatives of these materials becomes highly significant.

On the other side, recently, an array of diverse topological quantum states, including Weyl/Dirac semimetals\cite{xing2020surface,PhysRevB.84.220504,zhou2016pressure,schimmel2023high,jia2019superconductivity,PhysRevB.103.L081402} and nodal-line semimetals\cite{PhysRevB.96.165123,muechler2019superconducting,PhysRevB.103.L161109} have been experimentally observed in several superconducting systems such as ZrTe$_5$, TaN-series, TaIrTe$_4$, SnSe and YPtBi, etc. What's more, the interaction between these emergent topological non-trivial electronic states and superconductivity has been recognized as a promising platform for the realization of exotic quasiparticles,  holding great promise for applications in high-speed electronics and topological quantum computing\cite{Zhang_2020,PhysRevB.106.214510,TPspACSomega2023,chen2020unconventional}.

Previous theoretical work has discussed the possibility of placing O-atom at any one of the $4e$ Wyckoff Position (WP), which agree that Pb$_{10}$(PO$_{4}$)$_{6}$O is a insulator\cite{lai2023first,si2023electronic}. The compound is non-magnetic and it also has no spin polarization properties\cite{hao2023first}.
However, strictly speaking, replacing the O atoms are 1/4 occupied with one of the four equivalent positions, would break the crystal¡¯s original high symmetry,
and they have found it plays a crucial role in modulating the electronic transport properties along the chain direction.
T. Kun et al. predicted that the Pb$_{10}$(PO$_{4}$)$_{6}$O$_{4}$, with fully occupied $4e$ positions, belongs to metal and they suggested that the introduction of oxygen vacancies would lead to the disappearance of the flat bands in the parent material\cite{tao2023cu}. While K. Ogawa et al. conducted investigations into the reaction thermodynamics of Pb$_{10}$(PO$_{4}$)$_{6}$O$_{x}$, where $x$ ranges from 0 to 4, thereby proposing potential synthesis conditions for varying oxygen concentrations\cite{ACSEnergyLetters2023}. Despite extensive research efforts\cite{lai2023first,si2023electronic,hao2023first,tao2023cu,cabezas2023theoretical,liu2023metallic,shen2023phase,liu2023different}, few attention have been payed to the Pb$_{10}$(PO$_{4}$)$_{6}$O$_{4}$, especially regarding its underlying topological semimetal state within the electronic structure and unexplored magnetic properties.

In this work, combining the first-principles calculations and low-energy effective models, we conduct a comprehensive investigation on the bandstructure, topological properties and magnetism of Pb$_{10}$(PO$_{4}$)$_{6}$O$_{4}$ with with oxygen fully occupied $4e$ positions. First, we reveal origin of quasi-one-dimensional flat-band around Fermi level by DFT calculation and tight-binding model. Moreover, via symmetry analysis and effective $k{\cdot}p$ models, we uncover the coexistence of multi-dimensional fermions around Fermi level, including 0D quadratic Dirac fermion at A point, 1D QNL/LNL along $\Gamma$A/HK path, 1D HNL on ALM$\Gamma$ plane, and 2D NS on $k_z$=$\pi$ plane. In addition, the oxygen molecules chains contribute a weak ferromagnetic order, thus Pb$_{10}$(PO$_{4}$)$_{6}$O$_{4}$ becomes rare semi-half-metal, simultaneously hosting 100$\%$ spin polarization and quadratic Dirac and HNL fermions at the Fermi level.


\begin{figure}[t]
\includegraphics[width=8.5 cm]{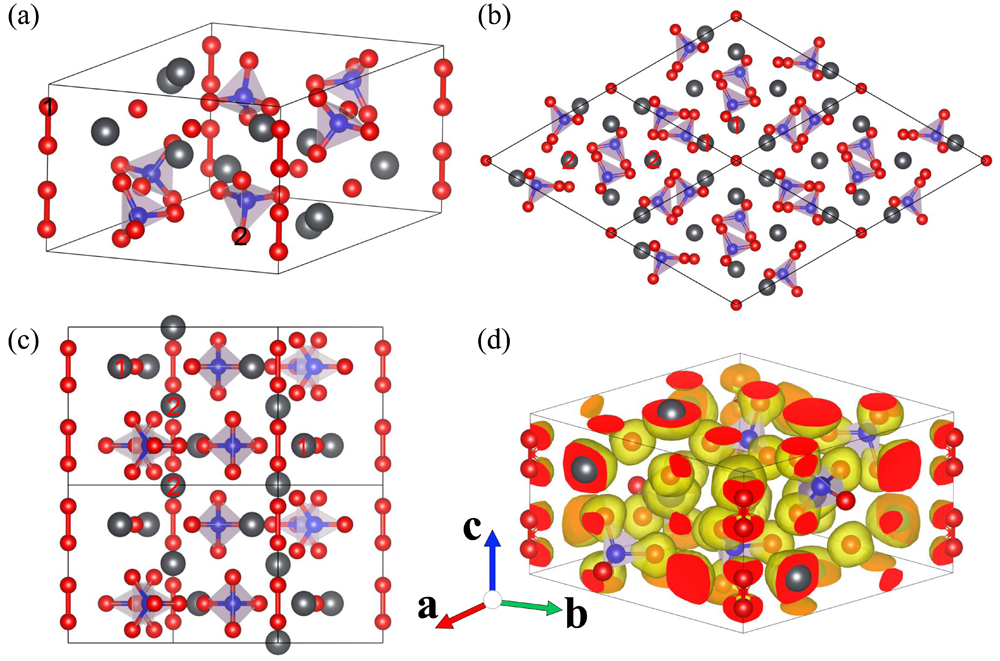}
\caption{Crystal structure of Pb$_{10}$(PO$_{4}$)$_{6}$O$_{4}$ with bird view (a), top view (b) and side view (c). (d) The ELF of Pb$_{10}$(PO$_{4}$)$_{6}$O$_{4}$.}\label{FIG1}
\end{figure}

\section{Computational Methods}
Structural optimization and electronic structure calculations were carried out by using Vienna $ab$-$initio$ simulation package (VASP)\cite{kresse1996efficiency,kresse1996efficient} with Perdew-Burke-Ernzerhof parameterized generalized gradient approximation (PBE-GGA) \cite{perdew1996generalized}. The energy cutoff of 500 eV and a \emph{k}-point mesh of 5$\times$5$\times$7 are chosen.
The ionic relaxations were performed  until the force on each atom was less than 0.01 eV\verb|/|\AA {} and convergence criterion for the self-consistent electronic minimization loop was set to 10$^{-6}$ eV. To analyze the electronic topological properties, the tight-binding (TB) Hamiltonian was constructed via MagneticTB code\cite{ZHANG2022108153}, and surface states were calculated using the iterative Green's function method as implemented in the WannierTools package\cite{wu2018wanniertools}.

\begin{figure*}[t]
\includegraphics[width=17.00 cm]{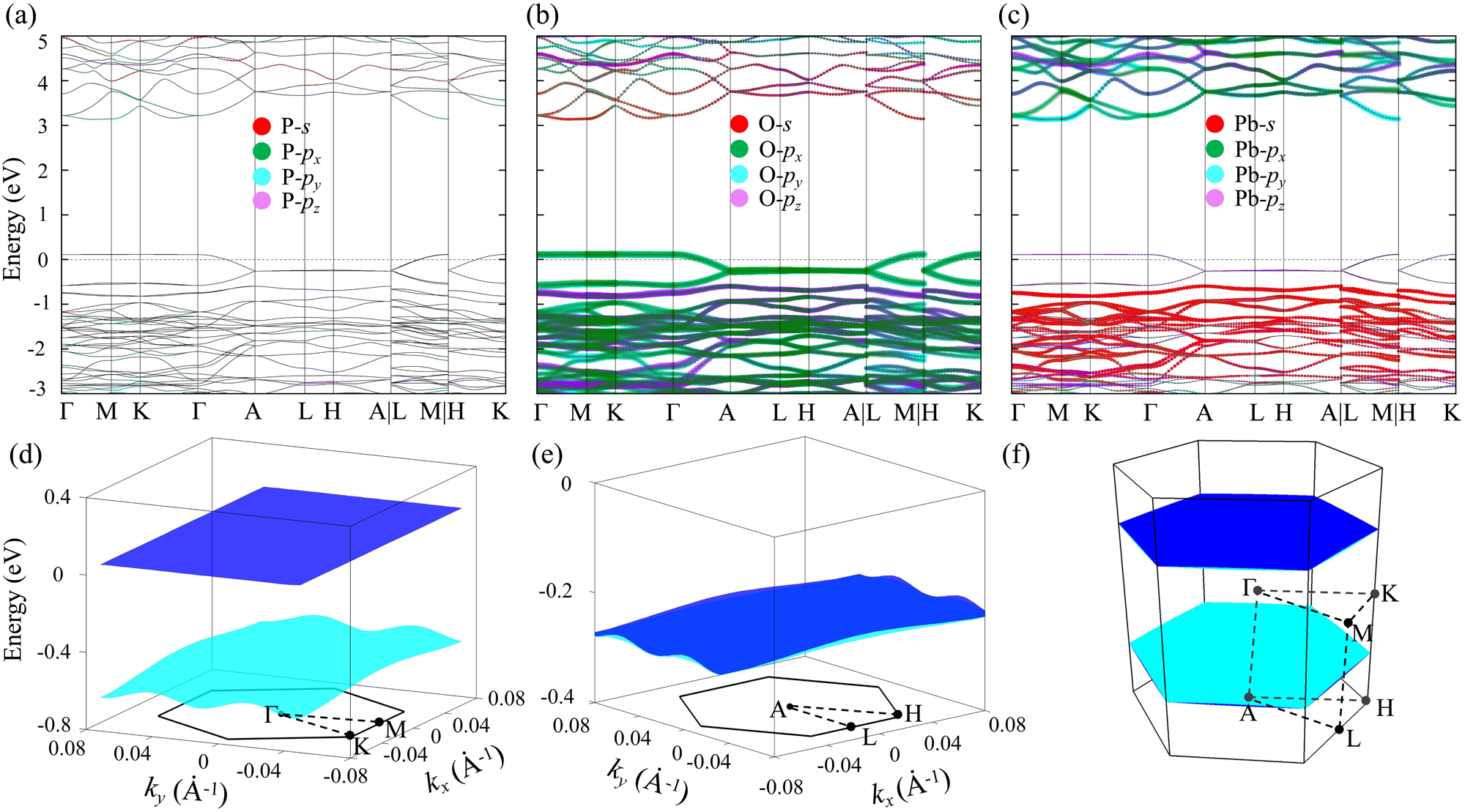}
\caption{(a)-(c) The band structures of Pb$_{10}$(PO$_{4}$)$_{6}$O$_{4}$ with different atomic orbital projections. The 3D view of four bands around Fermi level at $k_z=0$ plane in (d) and at $k_z=\pi$ plane in (e). (f) The Fermi surface of Pb$_{10}$(PO$_{4}$)$_{6}$O$_{4}$.}\label{FIG2}
\end{figure*}

\begin{figure*}[t]
\includegraphics[width=18.1 cm]{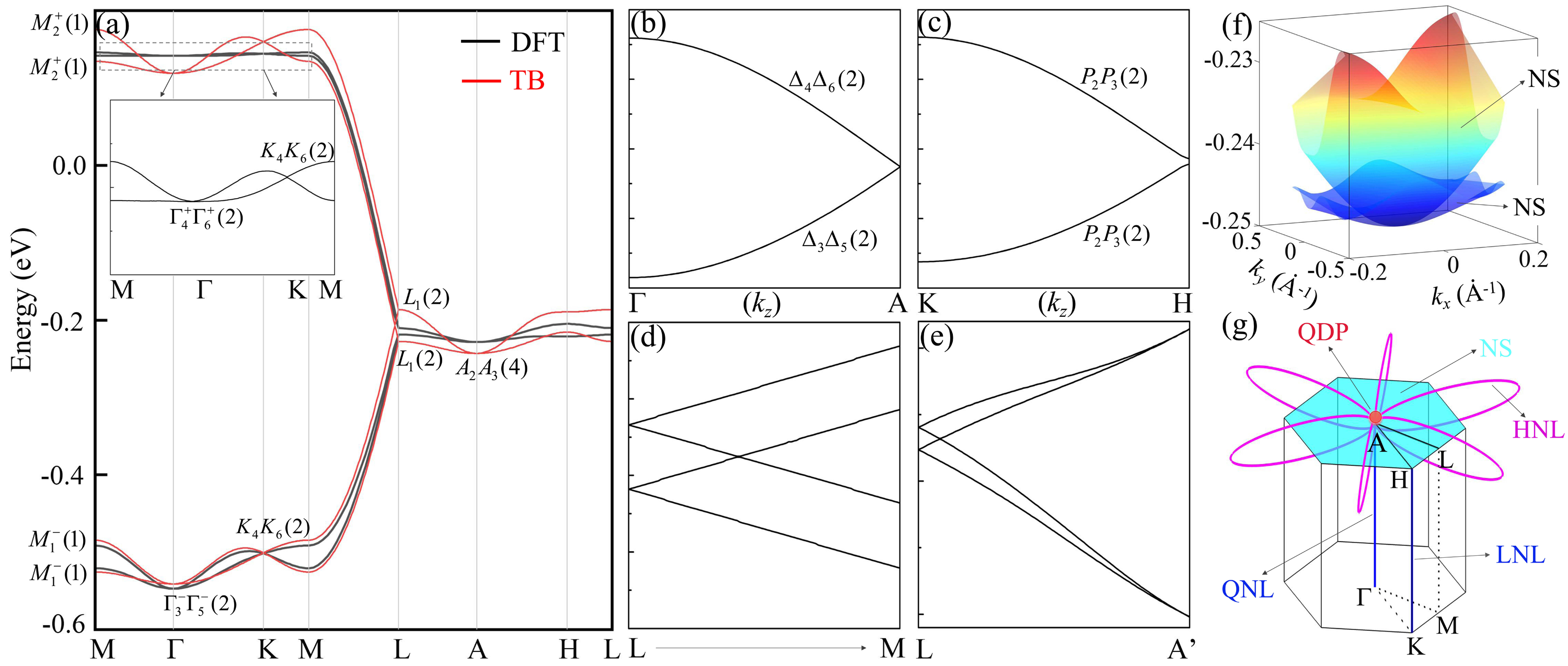}
\caption{(a) Comparison showing of four-bands of Pb$_{10}$(PO$_{4}$)$_{6}$O$_{4}$ around Fermi level by DFT and tight-binding (TB) model, where the fitting parameters of Hamiltonian are chosen as $e_1$=-0.224, $t_1$=0.165, $r_1$=0.008, $s_1$=-0.009 and $s_2$=-0.005. The
irreducible representations are marked for each band, where the superscript $\pm$ denotes the parity eigenvalues.
(b) Double degenerated band along $\Gamma$A path. (c) Double degenerated band along KH path. (d) Hourglass-like band dispersion along LM path, (e) Hourglass-like band dispersion along LA' path. The A' (0, 0, 0.48)located at $\Gamma$A path. (f) 3D view of quadratic Dirac point at A point on $k_z$=$\pi$ plane. (g) Schematically illustrations  for the location of multi-dimensional fermions in the Brillouin zone.}\label{FIG3}
\end{figure*}

\section{Results and anlysis}
\subsection{Crystal structure and chemical bonding of Pb$_{10}$(PO$_{4}$)$_{6}$O$_{4}$ }
The Pb$_{10}$(PO$_{4}$)$_{6}$O$_{4}$ has the same crystal structure with experimentally synthesized\cite{lee2023roomtemperature} Pb$_{10}$(PO$_{4}$)$_{6}$O with the space group of 176 (P6$_3$/m). The optimized lattice constants are $\bm{a}$=$\bm{b}$=10.151 {\AA} and $\bm{c}$=7.367 {\AA}, slightly larger than those of Pb$_{10}$(PO$_{4}$)$_{6}$O. The schematic structure is depicted in Fig.~\ref{FIG1}(a)-(c).
In Fig.~\ref{FIG1}(a), six phosphorus atoms are situated at the Wyckoff position (WP) of $6h$ (0.5912,0.9643, 0.75) and are bonded to four nearest neighboring oxygen atoms, forming six [PO4] tetrahedra.
The remaining four oxygen atoms uniformly occupy the $4e$ WP (0, 0, 0.6571), arranged in a row and neatly positioned on the rhombus of the unit cell.
As shown in Fig.~\ref{FIG1}(b), the ten lead atoms can be categorized into two groups: the first group is at the WP of $6h$ (0.2515,0.9882, 0.75), forming a regular hexagon centered around the oxygen atoms on the vertical edge,  the second group of lead atoms is located at $4f$ (1/3, 2/3, 0.4955), comprising two two layers as depicted in Fig.~\ref{FIG1}(c). These lead atoms are surrounded by those oxygen atoms from the [PO$_{4}$] tetrahedra, with each lead atom coordinated by six oxygen atoms, exhibiting bonding characteristics reminiscent of H-MoS$_2$.

The bonding configurations are vividly visualized through calculating the electron localization function (ELF)\cite{silvi1994classification} in Fig.~\ref{FIG1}(d). In the [PO4] tetrahedron, covalent bonds are observed between P and O atoms, with a bond length of 1.567 {\AA}. Notably, the P atom has no remaining lone pair electrons, indicating that all its valence electrons are engaged in bonding. Along the vertical edge of the oxygen atom chain, the intra- and inter-molecular distances are 1.721 {\AA} and 1.994 {\AA}, respectively. The ELF clearly reveals the presence of covalent bonds between the O atoms within the molecule. Additionally, a pair of lone pairs with the characteristics of $p_{x,y}$-${\pi}^*$ anti-bonding orbitals of O$_2$ molecular are clearly observed. Furthermore, Bader charge analysis indicates that Pb1 and Pb2 atoms lose 2.59 and 2.60 electrons, respectively, confirming the ionic bonding between Pb and [PO$_4$], which is consistent with the above analysis.

\subsection{Quasi-one-dimensional band structure of Pb$_{10}$(PO$_{4}$)$_{6}$O$_{4}$}
The electronic structure, as shown in Fig.~\ref{FIG2}(a)-(c), reveals that Pb$_{10}$(PO$_{4}$)$_{6}$O$_{4}$ exhibits metallic properties. Close to the Fermi level, we observe four bands that are well-separated from the rest bands, hosting pseudo band gaps above and below them.
Interestingly, in the $k_z=0$ and $k_z=\pi$ planes, these four bands exhibit strong anisotropic dispersion, where they display nearly flat dispersion along $k_x/k_y$ direction, known as ``flat bands", while showing strong dispersion along the $k_z$ direction.
To provide a clearer visualization of these flat bands, we present a 3D view of bands at $k_z=0$ and $\pi$ planes in Fig.~\ref{FIG2}(d) and (e). At $k_z=0$ plane, the two conduction bands (blue color) demonstrate minimal dispersion with a extremely narrow band width of only 5 meV, indicating a perfect flat band characteristic. Similarly, the two valence bands (cyan color) also exhibit weak dispersion with a band width of approximately 60 meV.
Moving to the $k_z=\pi$ plane, the four bands become in pairs, intertwined together. Here too, they display very weak dispersion with a band width of around 23 meV.

To delve deeper into the origins of these distinctive band structures, we conducted an orbital projection analysis. It's discovered that the conduction bands ($\geq$3 eV) mainly arise from the $p$-orbitals of Pb and the $s$-orbitals of O. Conversely, the valence bands around and below the Fermi level are primarily contributed by the $s$-orbitals of Pb and the $p$-orbitals of O. These can be understood as follows: Pb atom loses electron from $6p$ orbitals,  resulting in a substantial contribution to the conduction bands. Simultaneously, the $s$-orbitals of Pb are occupied by electrons, and this contributes to the formation of the valence bands below the Fermi level (-3 eV to -1 eV) as depicted in Fig.~\ref{FIG2}(c). O atom gains electrons, and hence, its $p$-orbitals mainly contribute to the below and around the Fermi level in Fig.~\ref{FIG2}(b).

Particularly, through charge density calculations in supplementary materials, we found that the four energy bands near the Fermi level primarily originate from the
anti-bonding molecular orbital of the oxygen at $4e$ WP positions:
\begin{equation} \label{kpf0}
|1, \pi_{p_x}^*>, |1, \pi_{p_y}^*>, |2, \pi_{p_x}^*>, |2, \pi_{p_y}^*>,
\end{equation}
the 1 and 2 refers two O$_2$ molecular, the $\pi_{p_{x,y}}^*$ stands for the anti $\pi$-bond formed by $p_{x,y}$-orbital of O-atoms. Along $z$ direction, the Hamiltonian is written as,
\begin{equation} \label{kpf0}
\mathcal{H}(k_z)=e_0 \tau_0 \sigma_0+t[(1+cos(k_z)\tau_x)+sin(k_z)\tau_y ]\sigma_x.
\end{equation}
The $\bm{\sigma}$ and $\bm{\tau}$ represent intra- and inter-molecule degrees of freedom while the $e_0$ is the molecular orbital energy level and the $t$ is the coupling strength between two molecules. It gives energy spectrum along $k_z$ as $E_{\pm}=e_0 \pm 2tcos(k_z/2)$ with a gap of $2t$ at $k_z$=0. As $k_z$ increases, these bands gradually approach each other, eventually crossing at $k_z$=$\pi$, in consistent with the DFT results in Fig.~\ref{FIG3}(b).

Previous structural analysis already reveals that these four oxygen atoms form an effectively one-dimensional oxygen molecule chain along the $z$ direction, with a large intermolecular distance (9.865 {\AA}), resulting in very weak in-plane coupling and
non-dispersive flat band. As shown in Fig.~\ref{FIG2}(f), this
quasi-one-dimensional electronic structure is also reflected from the Fermi surface, where the Fermi surface of the
system consists of two parallel planes, providing perfectly
Fermi surface nesting with a vector of (0, 0, 0.517).

\subsection{Multi-dimensional fermions embedded within the flat band of Pb$_{10}$(PO$_{4}$)$_{6}$O$_{4}$}

Because electrons near the Fermi level play a dominant role in transport properties, we will conduct an in-depth analysis of the four energy bands in close proximity to the Fermi level through through first-principles calculations, symmetry analysis, and low-energy effective models. A magnified view of the band structure around the Fermi level is depicted in Fig.~\ref{FIG3}(a).
Firstly, within the $k_z=0$ plane, these four energy bands can be categorized into two distinct groups. The lower-energy bands represent bonding states between oxygen molecules, while the higher-energy bands correspond to anti-bonding states. The separation between these two sets of bands is dictated by the strength of coupling between molecules along the $z$-direction, whereas the dispersion within each set is influenced by the in-plane coupling between O$_2$ molecules. Despite the relatively weak dispersion observed in the in-plane bands, meticulous calculations unveil that the two valence bands mutually interact within the energy range of -0.6 eV to -0.5 eV, resulting in a double degeneracy both at $\Gamma$ and K points. Notably, at the $\Gamma$ point, there is quadratic dispersion in the $k_{x}$-$k_{y}$ plane, while at K point, a linear dispersion is evident. Upon closer examination of the two conduction bands, a similar double degeneracy is observed at both the $\Gamma$ and K points.

Even more intriguingly, these instances of 2-fold degeneracy are not isolated. Actually, they persist consistently along the $k_z$ direction, resulting in the creation of a quadratic nodal-line (QNL) \cite{YuPhysRevB.99.121106} along the $\Gamma$A path and a linear nodal-line (LNL) along the KH path, as illustrated in Fig.~\ref{FIG3}(b) and~\ref{FIG3}(c). As $k_z$ evolves, the energy of the conduction bands decreases while the energy of the valence bands increases until they converge on the $k_z=\pi$ plane. In fact, due to the nonsymmorphic symmetry of 176 space group these bands will maintain a twofold degeneracy throughout the entire $k_z=\pi$ plane, which consequently gives rise to the formation of a 2D nodal surface (NS)\cite{PhysRevB.97.115125,PhysRevB.105.245152,nodalsurfaceNL2021}, as shown in Fig.~\ref{FIG3}(f).
Of particular interest is the A point, where the two doubly nodal surface intersect, forming a fourfold degenerate Dirac point with quadratic dispersion in the $k_x$-$k_y$ plane, namely forming a quadratic Dirac point \cite{PhysRevB.101.205134} (QDP).

Additionally, on the high-symmetric A$\Gamma$ML plane in the Brillouin zone (BZ), there exists new nodal line structure centered at L point.
As illustrated in Fig.~\ref{FIG3}(f), along LM and LA' paths, the 2-fold degenerate conduction and valence band are lift and inter-exchange partners, leading to a hourglass-like dispersion relation, creating new double degeneracy point between middle two-bands. Due to coexistence of space inversion ($\mathcal{P}$) and time reversal symmetry ($\mathcal{T}$) these crossing points are not isolated, they form a continuous closed curve in the momentum space. Indeed, it gives rise to a hourglass nodal-line (HNL)\cite{PhysRevB.96.155206,PhysRevB.104.L060301} centered at L point and linked to A point as schematically shown in Fig.~\ref{FIG3}(g). Considering the $\widetilde{\mathcal{C}}_{6z}$ symmetry, there will be six equivalent HNL which are linked to the central Dirac point, extending along AL direction, forming a nodal-network structure\cite{PhysRevB.93.201104,PhysRevB.98.075146,PhysRevLett.120.026402} in the extended BZ. As a whole, multi-dimensional fermions including 0D QDP at A point, 1D QNL/LNL along $\Gamma$A/KH path, 1D HNL lying on A$\Gamma$ML plane and 2D NS on $k_z=\pi$ plane are clearly demonstrated in Pb$_{10}$(PO$_{4}$)$_{6}$O$_{4}$.

\subsection{ The tight-binding model and low energy effective $\textbf{\emph{k}}$$\cdot$$\textbf{\emph{p}}$ model of Pb$_{10}$(PO$_{4}$)$_{6}$O$_{4}$}
To unravel the origin of multi-dimensional fermions in Pb$_{10}$(PO$_{4}$)$_{6}$O$_{4}$, we construct effective models through symmetry analysis.
The lattice of Pb$_{10}$(PO$_{4}$)$_{6}$O$_{4}$ belong to a nonsymmorphic space group $P6_{3}/m$ (No. 176) which hosts various types of symmetry operators such as rotation $(\mathcal{C}_{3z})$, spatial-inversion ($\mathcal{P}$), mirror ($\mathcal{M}_z$) and screw-rotation $(\widetilde{\mathcal{C}}_{2z})$.
Based on the first-principles calculations, we can find that appearance of the four-fold Dirac point at the $A$ arises from the essential band degeneracy, which corresponds to the 4D irreducible representations (IRRs) of $A_{2}A_{3}$ {[}see Fig.~\ref{FIG3}(a){]}.  Under the basis of $A_{2}A_{3}$, the matrix representations of the generators can be given as
\begin{eqnarray}
\widetilde{\mathcal{C}}_{6z}	=	\frac{\sqrt{3}}{2}\Gamma_{0,3}+\frac{i}{2}\Gamma_{3,3},\thinspace\thinspace\thinspace\mathcal{P}=\Gamma_{0,1},\thinspace\thinspace\thinspace\mathcal{T}=\Gamma_{1,0}\mathcal{K}
\end{eqnarray}
with $\Gamma_{i,j}=\sigma_{i}\otimes\sigma_{j}$, $\sigma_{0}$ as the $2\times2$ identity matrix and $\sigma_{i}$ $(i=1,2,3)$ as the Pauli matrix. $\mathcal{K}$ is a complex conjugate operator. With the standard approach~\cite{zhang2023magnetickp}, the effective Hamiltonian around the $A$ point retained to the leading order reads
\begin{eqnarray}
\mathcal{H}_{A}	&=&	\left[c_{1}+c_{2}(k_{x}^{2}+k_{y}^{2})\right]\Gamma_{0,0}+c_{3}\Gamma_{3,3}k_{z} \nonumber \\
		&+&\left(\alpha_{1}\Gamma_{+,1}k_{-}^{2}+\alpha_{2}\Gamma_{+,0}k_{+}k_{z}+h.c.\right).
\label{hamA}
\end{eqnarray}
Here, $\Gamma_{\pm,i}=\sigma_{\pm}\otimes\sigma_{i}$ with $\sigma_{\pm}=(\sigma_{1}\pm i\sigma_{2})/2$, $k_{\pm}=k_{x}\pm ik_{y}$.  $c_{i=1,2,3}$ is a real parameter, and $\alpha_{i}$ are complex parameters. Clearly, the obtained Hamiltonian (\ref{hamA}) exhibits linear band splitting along the $k_{z}$ direction and quadratic dispersion in $k_{x}$-$k_{y}$ plane, thus describes a QDP. Moreover, the QDP hosts a toplogical charge $\mathcal{C}$=0 due to the presence of $\mathcal{P}\mathcal{T}$ symmetry.

Besides the 0D Dirac point, one can observe a doubly degenerate nodal line along the $\Gamma$A path in Fig.~\ref{FIG3}(b). The little group of the $\Gamma$A path is $\mathcal{C}_{6}$, which is an abelian group without any 2D IRRs. Fortunately, there exists an anti-unitary operator $\mathcal{P}\mathcal{T}$ in the magnetic co-little group of $\Gamma$A, which can stick two 1D IRRs together to form a 2D IRRs $\Delta_{3}\Delta_{5}$ or $\Delta_{4}\Delta_{6}$ {[}see Fig.~\ref{FIG3}(b){]}, and thus protects a nodal line existence. We take $\Delta_{3}\Delta_{5}$ as an example, the matrix representation of symmetry generators can be expressed as
\begin{eqnarray}
\widetilde{\mathcal{C}}_{6z}	=-\frac{1}{2}\sigma_{0}+\frac{\sqrt{3}i}{2}	\sigma_{3},\thinspace\thinspace\thinspace\thinspace\mathcal{PT}=\sigma_{1}\mathcal{K}. \label{generator}
\end{eqnarray}
Under the symmetry constraints, the effective Hamiltonian around the $\Gamma$A path could be derived as
\begin{eqnarray}
\mathcal{H}_{{\Gamma}A}	&=&	\left[c_{1}+c_{2}k_{z}+c_{3}(k_{x}^{2}+k_{y}^{2})\right]\sigma_{0} \nonumber \\
		&+&\left(\alpha_{1}\sigma_{+}k_{+}^{2}+h.c.\right),
\label{hamGA}
\end{eqnarray}
where $c_{i}$ are real parameters, and $\alpha_{1}$ is a complex parameters. The Hamiltonian (\ref{hamGA}) describes a QNL located at the $\Gamma$A path, which exhibits linear dispersion along $k_{z}$ and quadratic dispersion normal to nodal line. For the other 2D IRRs $\Delta_{4}\Delta_{6}$ in the co-little group of $\Gamma$A, one can check that its effective Hamiltonian shares the same form as Eq. (\ref{hamGA}), which also describes a QNL.
Besides, applying above analysis to the double degenerate nodal line along KH path with IRRs $P_{2}P_{3}$ {[}see Fig. \ref{FIG3}(c){]}, the corresponding effective Hamiltonian is written as,
\begin{eqnarray}
H_{KH}	&=&	\left(c_{1}+c_{2}k_{z}\right)\sigma_{0}+\left(c_{3}k_{x}+c_{4}k_{y}\right)\sigma_{1} \nonumber \\
	&+&	\left(c_{3}k_{y}-c_{4}k_{x}\right)\sigma_{2}
\label{hamHK}
\end{eqnarray}
which describes a LNL as expected.

In addition, there also exists a NS with 2D manifold of reciprocal space due to the presence of twofold screw-rotational symmetry $\widetilde{\mathcal{C}}_{2z}=\left\{ \mathcal{C}_{2z}|0,0,\frac{1}{2}\right\}$ in Pb$_{10}$(PO$_{4}$)$_{6}$O$_{4}$. It is worth noting that a combination of twofold screw-rotational symmetry and and time-reversal symmetry $\mathcal{T}$ could guarantee the appearance of NS on boundary of the Brillouin zone (BZ)~\cite{whm2016PhysRevB.93.085427}. Specifically, one can find that $(\widetilde{\mathcal{C}}_{2z})^{2}=e^{ik_{z}}$, $\mathcal{T}^{2}=1$ for spinless case. Therefore, $\widetilde{\mathcal{C}}_{2z}\mathcal{T}$ is anti-unitary operator and satisfies $(\widetilde{\mathcal{C}}_{2z}\mathcal{T})^{2}=-1$ on the $k_{z}=\pi$ plane, which leads to Kramers-like degeneracy of NS. The result is also confirmed by our DFT calculations.

Since the multi-dimensional fermions including QDP, QNL/LNL, HNL and NS all originate from the isolated four bands near the Fermi surface, a simple four-band lattice model is possible to be established, and can be used for further investigation of the physical properties related to Pb$_{10}$(PO$_{4}$)$_{6}$O$_{4}$. Based on the first-principles calculations, we identify the IRRs of four bands near the Fermi level belong to a single elementary band representations (EBRs) $^{1}E^{''}{}^{2}E^{''}@2a$ of SG 176 in topological quantum chemistry~\cite{gao2021irvsp,bradlyn2017topological,cano2018building}. Therefore, we can construct a four-band lattice model by setting two basis orbitals $\left(\left|p_{x}\right\rangle ,\left|p_{y}\right\rangle \right)$ on $2a$ Wyckoff position in terms of the relationship between the site-symmetry of the Wyckoff position and the EBRs. The matrix representation of generators of SG $176$ are given by
\begin{eqnarray}
D(\mathcal{C}_{3z}) & = & \frac{1}{2}\Gamma_{0,0}-\frac{\sqrt{3}i}{2}\Gamma_{0,2},\thinspace\thinspace\thinspace D(\widetilde{\mathcal{C}}_{2z})=-\Gamma_{1,0}\\
D(\mathcal{P}) & = & -\Gamma_{1,0},\thinspace\thinspace\thinspace D(\mathcal{T})=\Gamma_{0,0}\mathcal{K}.
\end{eqnarray}
Resultantly, the symmetry allowed lattice model may be written as
\begin{widetext}
\begin{eqnarray}
\mathcal{H}_{TB} & = & e_{1}+s_{1}\left[\mathrm{cos}\left(k_{x}\right)+\mathrm{cos}\left(k_{y}\right)+\mathrm{cos}\left(k_{x}+k_{y}\right)\right]+2r_{1}\mathrm{cos}\left(k_{z}\right)\nonumber \\
 & + & \left[\left(\sqrt{3}s_{1}-s_{2}\right)\mathrm{cos}\left(k_{x}\right)+2s_{2}\mathrm{cos}\left(k_{y}\right)-\left(\sqrt{3}s_{1}+s_{2}\right)\mathrm{cos}\left(k_{x}+k_{y}\right)\right]\frac{\Gamma_{0,1}}{2} \nonumber \\
 & + & \left[2s_{1}\mathrm{cos}\left(k_{y}\right)-\left(\sqrt{3}s_{2}+s_{1}\right)\mathrm{cos}\left(k_{x}\right)+\left(\sqrt{3}s_{2}-s_{1}\right)\mathrm{cos}\left(k_{x}+k_{y}\right)\right]\frac{\Gamma_{0,3}}{2} \nonumber \\
 & + & 2t_{1}\mathrm{cos}\left(\frac{k_{z}}{2}\right)\Gamma_{1,0}+s_{2}\left[\mathrm{sin}\left(k_{x}\right)+\mathrm{sin}\left(k_{y}\right)-\mathrm{sin}\left(k_{x}+k_{y}\right)\right]\Gamma_{3,2}.
\label{tb}
\end{eqnarray}
\end{widetext}
Here, $e_{1}$ represents the onsite energy, $t_1$ and $r_1$ correspond to the nearest neighbor and next neighbor intra-chain hopping along the $z$-direction, respectively. Additionally, $s_i$ denotes the inter-chain coupling. In Fig.~\ref{FIG3}(a), we plot the electronic band structure of Eq. (\ref{tb}), one can observe the established lattice model indeed captures the main feature of Pb$_{10}$(PO$_{4}$)$_{6}$O$_{4}$ material with multi-dimensional Fermions.

\begin{figure}[t]
\includegraphics[width=8.6 cm]{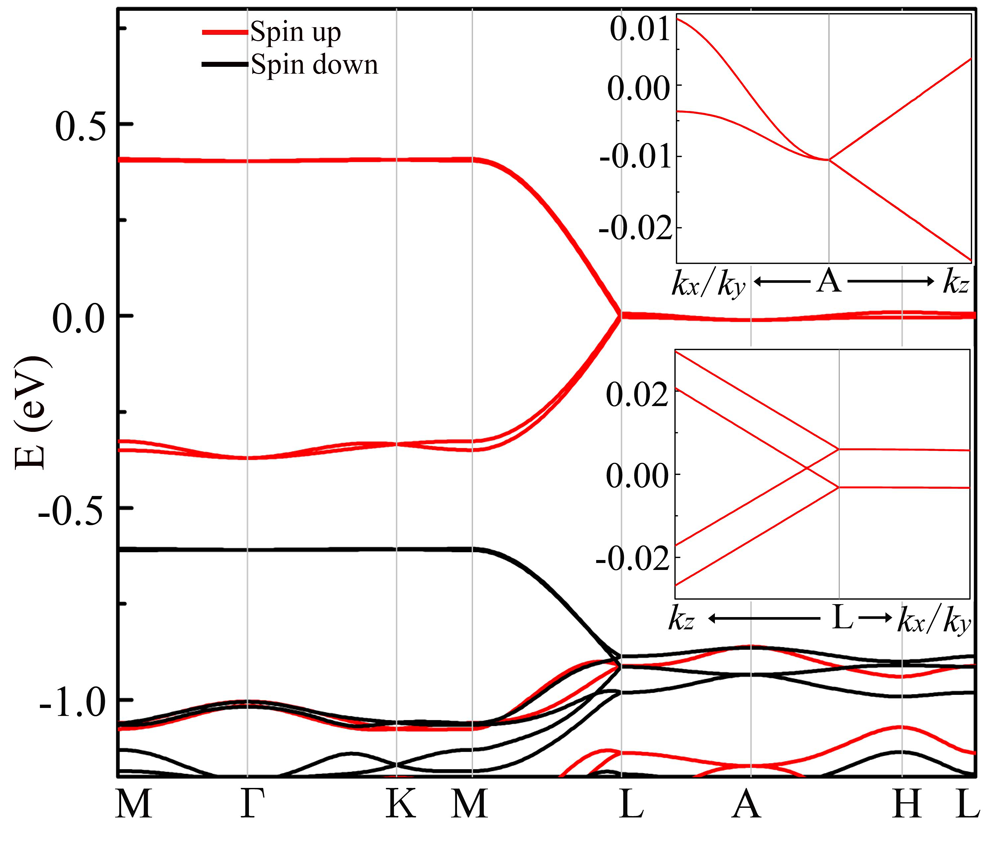}
\caption{Spin polarized band structure of Pb$_{10}$(PO$_{4}$)$_{6}$O$_{4}$. Insets are the enlarged bands around A and L points.}\label{FIG4}
\end{figure}

\section{semi-half-metal state of Pb$_{10}$(PO$_{4}$)$_{6}$O$_{4}$ under ferromagnetic ground state}
Building on our previous analysis, the low energy electron states of Pb$_{10}$(PO$_{4}$)$_{6}$O$_{4}$ are mainly come from the $4e$ O-atoms. These four oxygen atoms can be viewed as two pairs of interacting oxygen molecules.
Given the inherent paramagnetism of oxygen molecule (2.0 $\mu_B$) due to its half-filled 2-fold $\pi^{*}$ bond, we further explore the underlying magnetic properties of Pb$_{10}$(PO$_{4}$)$_{6}$O$_{4}$. The spin polarized calculation is carrier out and we found the system possesses a total magnetic moment of 2.0 $\mu_B$, primarily originating from O-atom at $4e$. The magnetic moment of O-atom in Pb$_{10}$(PO$_{4}$)$_{6}$O$_{4}$ is only half of that of isolated oxygen molecule. The reduction in the magnetic moment could be interpreted from two perspective. On the one hand, these oxygen atoms have gained electrons (0.4 $e$) from Pb atoms, leading to electron occupancy exceeding half-filling. On the other hand, there is electron hopping between two O$_2$ molecules, weaken the electron correlation effect. Both of these factors contribute to a reduction in the magnetic moment of the oxygen atoms compared to isolated O$_2$ molecules.

Moreover, by constructing intra-molecular and inter-molecular anti-ferromagnetic configurations, we discovered that the ferromagnetic configuration along the $z$ direction is the ground states (with 261 meV lower than the non-magnetic state and 217 meV lower than the anti-ferromagnetic state).
In Fig.~\ref{FIG4}, we present the spin-polarized energy bands, revealing a spin splitting that leads one channel to become a semiconductor while the other spin channel crosses the Fermi level. This configuration is known as a half-metal state.
What's particularly intriguing is the presence of completely spin-polarized flat bands at $k_z$=0 plane and topological semimetal states, including QDP at A point and the HNL around surrounding L point, all located at the Fermi energy level. This confirms Pb$_{10}$(PO$_{4}$)$_{6}$O$_{4}$ belongs to the rare semi-half-metal, which may exhibit immensely strong correlation effects and topological transport properties.
\section{Conclusions}
In summary, our theoretical investigation has unveiled the presence of flat bands and novel topological semimetal states in the LK-99 derivative. We have elucidated that the origin of the nearly flat dispersion around the Fermi level primarily stems from a quasi 1D O-atom chain. Furthermore, the unique arrangement of O-atoms with nonsymmorphic symmetry plays a crucial role in generating a 4-fold QDP at the A point, accompanied by the presence of highly interesting HNL fermions. To comprehensively capture the topological features of diverse multi-dimensional fermions embedded in Pb10(PO4)6O4, we have developed both local low-energy k¡¤p models and a global four-band tight-binding model.
Our research not only unveils a promising material but also opens up avenues for exploring the interplay between nontrivial electronic states, magnetism, and superconductivity within the Pb$_{10}$(PO$_{4}$)$_{6}$O$_{4}$.

\begin{acknowledgements}
This work is supported by the National Natural Science Foundation of China (NSFC, Grants No. 12304086, No. 12204330), and Dr. B. Fu also the Sichuan Normal University for financial support (Grant No. 341829001).
\end{acknowledgements}

\end{document}